\documentclass[%
jmp,%
 sd,%
amsmath,amssymb,
preprint,%
]{revtex4-1}

\usepackage{graphics}
\usepackage{dcolumn}
\usepackage{bm}

\begin{document}


\title[]{Quantum Propagator Dynamics of a Harmonic Oscillator in a Multimode Harmonic Oscillators Environment using White Noise Functional Analysis} 

\author{B. M. Butanas Jr.}
  \altaffiliation[]{bmbutanas@upd.edu.ph}
  \affiliation{%
Research Cluster for Quantum and Biological Systems, Theoretical Physics Group, National Institute of Physics, College of Science, University of the Philippines, Diliman, Quezon City, 1101, Philippines
}%
  \affiliation{%
Department of Physics, College of Arts and Sciences, Central Mindanao University, Musuan, Maramag, Bukidnon, 8710 Philippines
}%

\author{R. C. F. Caballar}
 \altaffiliation[]{rfcaballar@up.edu.ph}
\affiliation{%
Research Cluster for Quantum and Biological Systems, Theoretical Physics Group, National Institute of Physics, College of Science, University of the Philippines, Diliman, Quezon City, 1101, Philippines
}%

\date{\today}

\begin{abstract}
White noise analysis is used to derive the propagator of an open quantum system consisting of a harmonic oscillator which is coupled to an environment consisting of N multimode harmonic oscillators. The quantum propagators are obtained after solving for the normal modes of the system-environment interaction in order to decouple the coordinates in the Lagrangian describing the dynamics of the system, the environment and their interaction with each other. The decoupled Lagrangian is then used in the path integral corresponding to the propagator of the system, with the path integral evaluated using white noise analysis. The resulting propagator is then found to consist of a product of N simple harmonic oscillator propagators. N-2 of these propagators correspond to the degenerate normal mode frequencies of the system-environment interaction, while the other 2 correspond to the non-degenerate normal mode frequencies. This method of deriving the propagators greatly simplifies the task of mathematically describing the dynamics of the open quantum system. The ease of use of this method suggests that, together with its inherent mathematical rigor, white noise analysis can be a powerful tool in analyzing the dynamics of an open quantum system.
\end{abstract}


\maketitle


\section{\label{sec:intro}Introduction}

System-environment interactions in quantum mechanics have recently become a topic of intense study, enabling us to better understand dissipation and decoherence in quantum mechanics \cite{i.a}-\cite{i.n}. One physical model that has been used to study system-environment interactions in quantum mechanics is the Caldeira-Leggett (CL) model \cite{i.a, a.a}. The CL model describes a quantum system with an arbitrary potential interacting with an environment modeled as an infinite number of harmonic oscillators \cite{i.a}. The influence-functional method of Feynman and Vernon \cite{a.a2} was used to analyze this model. In doing so, the master equation describing the dissipative dynamics of the system was obtained. The system-bath interactions in this model enables us to describe dissipation phenomena in solid state physics, quantum tunneling and quantum computing\cite{a.a1, a.a1a}. 

The exact master equation for the CL model with the system composed of a single harmonic oscillator interacting with an environment of an infinite number of harmonic oscillators was solved using the influence functional \cite{i.b}, Wigner function \cite{a.a3} and quantum trajectories method \cite{a.a4}. The propagator for this particular CL model was solved in Refs. \cite{i.e, a.b}. These studies can be extended and generalized towards the analysis of a system composed of N harmonic oscillators coupled to an environment modeled as an ensemble of harmonic oscillators. Such an extension and generalization is vital in understanding macroscopic quantum phenomena such as decoherence since any quantum system, and the environment with which it is interacting, can be decomposed into a number of components which are modeled as harmonic oscillators.  

In this paper, we consider a system composed of a single harmonic oscillator interacting with an environment composed of an ensemble of harmonic oscillators. We will use a method different from those used in Refs. \cite{i.b, i.e, a.a, a.b} to derive the quantum propagator of the system. This method makes use of white noise analysis invented by Hida \cite{a.c}. As compared to the influence functional method which is considered to be mathematically ill-defined due to the presence of the Lebesgue measure, white noise analysis is a mathematically well-defined method, and is a powerful tool in evaluating the Feynman path integral \cite{a.d}. It is our motivation here to show the promise of white noise analysis in evaluating propagators for open quantum systems.     

This paper is organized as follows. The first section will present the system and the bath, together with their corresponding Hamiltonians, considered in this study. The second and third sections will discuss some basics on white noise analysis and the recasting of Feynman path integral in the context of white noise analysis following Ref. \cite{a.d}. Finally, the fourth section will present the application of white noise analysis to derive the quantum propagator, before we conclude with a short summary. 

\section{Mathematical Preliminaries}
In this section, we review the fundamentals of white noise analysis, which is the mathematical tool that will be used in evaluating the Feynman path integral arising in the quantum propagators describing the dynamics of the system. 

\subsection{Harmonic Oscillator in an Environment of a Finite Number of Harmonic Oscillators}
\label{coupled}

\begin{figure}
\includegraphics{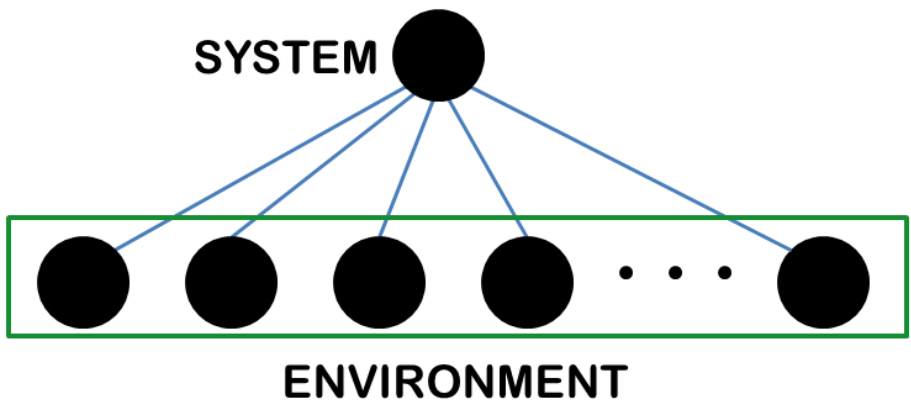}
\caption{\label{fig:system} Schematic diagram of the simple harmonic oscillator system coupled to an environment of $N$ harmonic oscillators, with the coupling of the system to each individual oscillator in the environment represented in the diagram by solid blue lines.}
\end{figure}

The Hamiltonians of the system $H_{S}$, which is a simple harmonic oscillator, the environment $H_{B}$, consisting of $N$ harmonic oscillators (see Fig. (\ref{fig:system})); and the interaction between the system and the environment $H_{SB}$, are defined respectively as follows: \cite{a.a} 
\begin{eqnarray}
H_{S} &=& \frac{p^2}{2m} + \frac{1}{2}m\omega^2 x^2, 
\label{coupled hs}
\\
H_{B} &=& \sum_{n=1}^{N}\left(\frac{p_{q_{n}}^2}{2m} + \frac{1}{2}m\omega_{n}^2 q_{n}^2\right), 
\label{hb}
\\
H_{SB} &=& C x \sum_{n=1}^{N} q_{n},
\label{sb}
\end{eqnarray}
where $x, q_{n}$, $p, p_{q_{n}}$, $\omega_{n}$ and $\omega$, are the corresponding positions, momenta, and frequencies of the system and bath oscillators, while $C$ is the coupling constant of the system-environment interaction. 
Moreover, we assume that the environment coordinate is linearly coupled to the system, with the coupling described in terms of the Hamiltonian $H_{SB}$, given above. Thus, the total Hamiltonian can be written as 
\begin{equation}
H = H_{S} + H_{B} + H_{SB}.
\label{totalH}
\end{equation}

\subsection{White Noise Analysis Fundamentals \cite{a.d}}
\label{whitenoise}

Formally, a stochastic process like Brownian motion obeys the stochastic differential equation given by
\begin{equation}
dX = a(t,X) dt + b(t,X) dB(t),
\label{stochasticdifeqn}
\end{equation}
where $X$ describes the Brownian motion, $B(t)$ is the Wiener process, $a(t,X)$ and $[b(t,X)]^2$ are the drift and diffusion coefficients, respectively. One can rewrite this stochastic differential equation as a Langevin equation, which is
\begin{equation}
\dot{X} = a(t,X) + b(t,X) \omega(t),
\label{langevin}
\end{equation}
where $\dot{X}=\frac{dX}{dt}$ and $\omega(t)=\frac{dB(t)}{dt}$ is interpreted as the velocity of Brownian motion, and is called the Gaussian white noise. We can rewrite the Gaussian white noise in terms of Wiener's Brownian motion i.e. $B(t)=\int_{t_{o}}^{t}\omega(\tau) d\tau=\left\langle \omega, 1_{[t_{o},t)}\right\rangle$, where we define $\left\langle \omega, \xi \right\rangle \equiv \int_{t_{o}}^{t}\omega(\tau) \xi(\tau) d\tau$.

Now, a key feature of Hida's formulation is the treatment of the set $\omega(\tau)$ at different instants of time, $\left\{ \omega(\tau); t \in \Re \right\}$ as a continuum coordinate system. For the sum over all routes or histories in the path integral, paths starting from initial point $x_{o}$ and propagating in Brownian fluctuations are parametrized within the white noise framework as
\begin{equation}
x(t) = x_{o} + \int_{t_{o}}^{t}\omega(s) ds.
\label{parametizewithmemory}
\end{equation}
Eq. (\ref{parametizewithmemory}) shows how the value of $x(t)$ is affected by its history, or earlier values of the modulated white noise variable $\omega(s)$ as $s$ ranges from $t_{o}$ to $t$. 

A further key feature of white noise analysis is that it operates in the Gelfand triple \cite{b.a} $S \subset L^{2} \subset S^{*}$, linking the spaces of a Hida distribution $S^{*}$ and test function $S$ through a Hilbert space of square integrable functions $L^{2}$. Using Minlos' theorem we can formulate a Hida white noise space $(S^{*},B,\mu)$ where $\mu$ is the probability measure and $B$ is the $\sigma$-algebra generated on $S$, and define a characteristic functional $C(\xi)$ given by
\begin{equation}
C(\xi) = \int_{S^{*}}\exp \left( \left\langle \omega, \xi \right\rangle \right) d\mu(\omega) = \exp \left( -\frac{1}{2}\int \xi^2 d\tau \right),
\label{characteristic}
\end{equation}
where $\xi \in S$ and the white noise Gaussian measure $d\mu(\omega)$ is given by
\begin{equation}
d\mu(\omega) = N_{\omega}\exp \left( -\frac{1}{2}\int \omega^2(\tau) d\tau \right) d^{\infty}\omega,
\label{gaussianmeasure}
\end{equation}
with $N_{\omega}$ as a normalization constant. The exponential term in $d\mu(\omega)$ is responsible for the Gaussian fall-off of the propagator function.

Now, the evaluation of the Feynman integral in the context of white noise analysis is carried out by the evaluation of the Gaussian white noise measure $d\mu(\omega)$. There are two important Gaussian white noise measure evaluation methods; these are the use of $T$- and $S$-transforms. For the $T$-transform of a generalized white noise functional $\Phi(\omega)$ we have the form
\begin{equation}
T\Phi(\xi) = \int_{S^{*}}\exp \left( i\left\langle \omega, \xi \right\rangle \right) \Phi(\omega) d\mu(\omega),
\label{Ttransform}
\end{equation}
similar to that of an infinite-dimensional Gauss-Fourier transform. On the other hand, the $S$-transform is related to the $T$-transform as follows:
\begin{equation}
S\Phi(\xi) = C(\xi) T\Phi(-i\xi),
\label{Stransform}
\end{equation}
where $C(\xi)$ is the characteristic functional given in Eq. (\ref{characteristic}).

\subsection{Feynman Quantum Propagator as a White Noise Functional \cite{a.d}}
\label{feymanevaluation}

The propagator for the quantum mechanical oscillator, say in x-dimension, has the following form as derived by Feynman \cite{c.a}:
\begin{equation}
K(x,x_{o};\tau) = \int \exp \left( \frac{i}{\hbar}S \right) D[x],
\label{propa2}
\end{equation}
where $S$ is the classical action defined as $S=\int L ~dt$ with $L=\frac{1}{2}m\dot{x}^2 -V(x)$ as the Lagrangian of the system, and $D[x]$ is the infinite-dimensional Lebesgue measure. Eq. (\ref{propa2}) sums over all the possible paths taken by a system/particle from an initial point $x(t_{o}=0) = x_{o}$ to a final point $x(t) = x$. Now, to rewrite this using the white noise analysis approach \cite{b.a, a.d} we introduce the parametrization of the path given by
\begin{equation}
x(t) = x_{o} + \sqrt{\frac{\hbar}{m}}\int_{0}^{t}\omega(\tau)d\tau,
\label{parametrize}
\end{equation}
Then taking the derivative of Eq. (\ref{parametrize}), substituting it into Eq. (\ref{exponential}) and simplifying the resulting equation, we obtain the exponential expression in the right hand side of Eq. (\ref{propa2}) as
\begin{equation}
\exp\left(\frac{i}{\hbar}S\right)= \exp \left[ \frac{i}{2}\int_{0}^{t}\omega(\tau)^2 d\tau \right] \exp \left[ -\frac{i}{\hbar}\int_{0}^{\tau}V(x) d\tau \right].
\label{exponential}
\end{equation} 
On the other hand, evaluation of the Lebesgue measure $D[x]$ leads to an integration over the Gaussian white noise measure $d\mu(\omega)$ in the relation
\begin{equation}
D[x]=\lim_{N\rightarrow \infty}\prod^{N}(A_{j})\prod^{N-1}(dx_{j})=Nd^{\infty}x, 
\label{gaussianmeasure0}
\end{equation} 
with
\begin{equation}
Nd^{\infty}x \rightarrow Nd^{\infty}\omega = \exp \left[ \frac{1}{2}\int_{0}^{t}\omega(\tau)^2 d\tau \right] d\mu(\omega),
\label{gaussianmeasure}
\end{equation}
where $N$ is the normalization constant. However, the path parametrization of the Brownian motion in Eq. (\ref{parametrize}) shows that only the initial point $x_{o}$ is fixed while the final point is fluctuating. Thus, to fix the endpoint we use the Fourier decomposition of a Donsker delta function, $\delta(x(t)-x)$, defined as
\begin{equation}
\delta(x(t)-x)=\frac{1}{2\pi}\int_{-\infty}^{+\infty}\exp \left( i\lambda (x(t)-x) \right) d\lambda,
\label{donsker}
\end{equation}
such that at time $t$ the particle is located at $x$. Finally, with Eqs. (\ref{exponential}), (\ref{gaussianmeasure}) and (\ref{donsker}) we now write the Feynman propagator in the context of white noise analysis as
\begin{eqnarray}
K(x,x_{o};\tau)&=&N \int \exp \left[ \frac{i+1}{2}\int_{0}^{t}\omega(\tau)^2 d\tau \right] 
\label{whitenoisepropagator}
\\
&\times&\exp \left[ -\frac{i}{\hbar}\int_{0}^{\tau}V(x)d\tau \right] \delta(x(t)-x) d\mu(\omega).
\nonumber
\end{eqnarray}

\section{Results and Discussion}
We derive the propagator for the system as follows. We first solve for the eigenvalues which correspond to the normal modes of the entire system-environment interaction. This allows us to decouple the system and environment coordinates by determining the amplitudes of the eigenvectors. After the system and environment coordinates have been decoupled, we then proceed to rewrite the Lagrangian and evaluate the path integral corresponding to the propagator for the system. We evaluate the path integral using the white noise analysis, which gives us a mathematically rigorous yet straightforward way to compute for the propagator of the system. 

\subsection{Normal modes}
Consider again the system-environment interaction in Fig. (\ref{fig:system}) and by the method of small oscillations we define the coordinates 
\begin{equation}
\eta_{i}=x_{i}-x_{oi},
\end{equation}
where $\eta_{i}$ is the small oscillation displacement and $x_{oi}$ is the distance between the system and $i^{th}$ harmonic oscillator environment. Then the potential and kinetic energy for the system and environment can be defined as follows:
\begin{equation}
V = \frac{1}{2}C(\eta_{1}-\eta_{2})^2 + \frac{1}{2}C(\eta_{1}-\eta_{3})^2 + \frac{1}{2}C(\eta_{1}-\eta_{4})^2 + \cdots + \frac{1}{2}C(\eta_{1}-\eta_{N+1})^2,
\label{potential}
\end{equation}
and
\begin{equation}
T = \frac{1}{2}m \dot{\eta_{1}}^2 + \frac{1}{2}m\dot{\eta_{2}}^2 + \frac{1}{2}m\dot{\eta_{3}}^2 + \cdots + \frac{1}{2}m\dot{\eta_{N}}^2. 
\label{kinetic}
\end{equation}
In the expression for the potential energy, the coordinates $\eta_{i}$ are coupled with each other. In order to decouple them, we compute for the normal mode frequencies of the system. The normal mode frequencies are the eigenvalues of the characteristic equation defined as $\left|V-\omega^2 T \right|=\left|V-\lambda I \right|=0$ or
\[
0 =  \left[ 
{\begin{array}{ccc}
V_{11} - \lambda  & V_{12} & \cdots V_{1N} \\
 V_{21} &  V_{22} - \lambda  & \cdots V_{2N} \\
 \vdots & \vdots &  \vdots \\
V_{N1} & V_{N2} &  \cdots V_{NN}-\lambda 
\end{array}}
 \right] ,
\]
where $I$ is an $N \times N$ matrix, $\lambda = \omega^2 m$ and
\begin{equation}
V_{ij} = \frac{\partial^2 V}{\partial q_{i} \partial q_{j}}, \hspace{0.5in} T_{ij} = m \frac{\partial \vec{r}_{k}}{\partial q_{oi}} \frac{\partial \vec{r}_{k}}{\partial q_{oi}}, 
\label{potential2}
\end{equation}
with $q$'s as the generalized coordinates. Inserting Eq. (\ref{potential2}) into the characteristic equation yields,
\[
0 =  \left[ 
{\begin{array}{ccc}
NC - \lambda  & C & \cdots C \\
-C &  -C - \lambda  & \cdots 0 \\
 \vdots & \vdots &  \vdots \\
-C & 0 &  \cdots -C-\lambda 
\end{array}}
 \right] ,
\]
where $N$ has integer values greater than or equal to two (2). Solving for the eigenvalues of this matrix, we find that for $N\geq 4$, there are $N-2$ degenerate eigenvalues, all of which are equal to $-C$, and two nondegenerate eigenvalues, which are given for increasing values of $N$ in Table~\ref{tab:table}.
\begin{table}
\caption{\label{tab:table}List of nondegenerate eigenvalues corresponding to the normal modes from $N=2$ to $N=10$.}
\begin{ruledtabular}
\begin{tabular}{lc}
N & Non-degenerate\\
  & Eigenvalues\\
\hline
2 & $\frac{1}{2}C \pm \frac{\sqrt{5}}{2}C$ \\
3 & $-C $\\
  & $ C \pm \sqrt{2}C $\\
4 & $\frac{3}{2}C + \frac{\sqrt{13}}{2}C$\\
  & $\frac{3}{2}C - \frac{\sqrt{13}}{2}C$\\
5 & $2C + \sqrt{5}C$\\
  & $2C-\sqrt{5}C$\\
6 & $\frac{5}{2}C+\frac{\sqrt{29}}{2}C$\\
  & $\frac{5}{2}C-\frac{\sqrt{29}}{2}C$\\
7 & $3C+\sqrt{10}C$\\
  & $3C+\sqrt{10}C$\\
8 & $\frac{7}{2}C+\frac{\sqrt{53}}{2}C$\\
  & $\frac{7}{2}C-\frac{\sqrt{53}}{2}C$\\
9 & $4C+\sqrt{17}C$\\
  & $4C-\sqrt{17}C$\\
10 & $\frac{9}{2}C+\frac{\sqrt{85}}{2}C$\\
   & $\frac{9}{2}C-\frac{\sqrt{85}}{2}C$\\
\end{tabular}
\end{ruledtabular}
\end{table}

By solving for the normal modes, we find that not only do we decouple the coordinates $\eta_{i}$, but we also manage to decompose the dynamics of the system into two components. The first component of the dynamics, described by the $N-2$ degenerate eigenvalues with value $-C$, correspond to only two oscillators in the environment moving in sync with each other, with the rest of the environment, as well as the system, remaining at rest. The second component of the dynamics, described by the two nondegenerate eigenvalues, correspond to both the system and the environment moving in sync with each other. This result holds for any number $N\geq 4$ of oscillators in the environment.

\subsection{Evaluation of the Path Integral using White Noise Analysis for $N=4$ Case}
\label{N4}
To illustrate the concepts presented in the previous subsection, let us analyze the dynamics of a simple harmonic oscillator interacting with an environment composed of $N=4$ harmonic oscillators. From Table~\ref{tab:table}, we obtain the eigenvectors corresponding to the normal modes for $N=4$ given by
\[
\vec{A} =  \left[ 
{\begin{array}{cccc}
 0 & 0 & \frac{-3C}{-(\frac{3}{2}+\frac{\sqrt{13}}{2})C+4C} & \frac{-3C}{-(\frac{3}{2}-\frac{\sqrt{13}}{2})C+4C} \\ 
-1 &-1 & 1 & 1 \\
 0 & 1 & 1 & 1 \\
 1 & 0 & 1 & 1
\end{array}}
 \right]. 
\]
Furthermore, we note that in general we can deduce what particular elements are oscillating through the equation, $\eta_{i}=a_{ij}\exp(i\omega_{i}t)$, where $a_{ij}$ corresponds to the amplitude of a certain element on entire system-environment interaction and $\omega_{i}$ as the normal mode frequency.

\subsubsection{Propagator for the dynamics described by the first degenerate normal mode frequency}
\label{1steigen}

Having computed for the eigenvectors corresponding to the normal modes, we can now explore the dynamics for every normal mode (eigenvalue) and evaluate the propagator describing the dynamics using white noise path integration. Consider the first eigenvalue $-C$ whose eigenvector is given by the first column of $\vec{A}$. From the eigenvector, $x_{1}=x_{3}=0$ and $x_{2}=-x_{4}$, hence the total Hamiltonian given in Eq. (\ref{totalH}) becomes, 
\begin{equation}
H = \frac{p_{2}^2}{2m} + \frac{1}{2}m\omega^2 x_{2}^2 + \frac{p_{4}^2}{2m} + \frac{1}{2}m\omega^2 x_{4}^2, 
\label{h1}
\end{equation}
where $\omega_{2}=\omega_{4}=\omega = \sqrt{\frac{\lambda}{m}}$. Note that we replaced our coordinate notation from $\eta$ or $q $ to $x$ for simplicity, since they represent position coordinates. Then, we solve for the total Lagrangian using Hamilton's canonical transformations 
\begin{equation}
\dot{q}_{k} = \frac{\partial H}{\partial p_{k}}, \hspace{0.5cm} -p_{k} = \frac{\partial H}{\partial q_{k}},
\nonumber
\end{equation}
and the relation given by 
\begin{equation}
L = \sum_{k}p_{k}\dot{q}_{k} - H,
\nonumber
\end{equation}
which yields the total Lagrangian as 
\begin{eqnarray}
L=\frac{1}{2}m\dot{x}_{2}^2 - \frac{1}{2}m\omega^2{x}_{2}^2 + \frac{1}{2}m\dot{x}_{4}^2 - \frac{1}{2}m\omega^2{x}_{4}^2. 
\label{l2}
\end{eqnarray}
Notice that the Lagrangian is separable into 
\begin{eqnarray}
L_{1} &=& \frac{1}{2}m\dot{x}_{2}^2 - \frac{1}{2}m\omega^2 x_{2}^2,
\label{l1newest}
\\
L_{2} &=& \frac{1}{2}m\dot{x}_{4}^2 - \frac{1}{2}m\omega^2 x_{4}^2.
\label{l2newest}
\end{eqnarray}
Clearly, it is evident that the total Lagrangian is separable into propagators for two independent harmonic oscillators which enable us to smoothly evaluate the Feynman path integral in the context of white noise analysis. Moreover, the classical action can be written as $S = \int_{0}^{t}L_{1}d\tau + \int_{0}^{t}L_{2}d\tau \Rightarrow S_{1}+S_{2}$. Thus, the full propagator can be written as $K(x_{2},x_{4};x_{2o},x_{4o};\tau) =K_{x_{2}x_{4}}=K(x_{2},x_{2o};\tau) K(x_{4},x_{4o};\tau)$ where 
\begin{eqnarray}
K(x_{2},x_{2o};\tau)=K_{x_{2}}=\int_{}^{}\exp \left[ \frac{i}{\hbar}S_{2} \right] D[x_{2}],
\label{k1}
\\ 
K(x_{4},x_{4o};\tau)=K_{x_{4}}=\int_{}^{}\exp \left[ \frac{i}{\hbar}S_{4} \right] D[x_{4}].
\label{k2}
\end{eqnarray}
Hence, we have shown that the full propagator is likewise separable. We then proceed with the evaluation of each individual propagator using white noise analysis.

\subsubsection*{The evaluation of the propagator corresponding to the Lagrangian $L_{1}$}

We substitute Eq. (\ref{l1newest}) and the classical action into Eq. (\ref{whitenoisepropagator}) and in doing so, we obtain the following propagator:
\begin{eqnarray}
K_{x_{2}}&=&N \int \exp \left[ \frac{i+1}{2}\int_{0}^{t}\omega(\tau)^2 d\tau \right] 
\label{whitenoisepropagatorK1}
\\
& & \times \exp \left[ -\frac{i}{\hbar}\int_{0}^{t}S_{V}(x_{2})d\tau \right] \delta(x(t)-x_{2}) d\mu(\omega),
\nonumber
\end{eqnarray}
where $S_{V}(x_{2})$ is just a term for the effective action of the harmonic oscillator potential. We parametrize the Donsker-delta function in Eq. (\ref{donsker}) as follows:
\begin{eqnarray}
\delta(x(t)-x_{2})=\frac{1}{2\pi}\int_{-\infty}^{+\infty}\exp \left[ i\lambda (x_{2o}-x_{2}) \right] \exp \left[ i\lambda \int_{0}^{t}\omega(\tau)d\tau \right] d\lambda, 
\label{donsker2}
\end{eqnarray}
We also parametrize the second exponential expression, which contains the potential, at the right hand side of Eq. (\ref{whitenoisepropagatorK1}), which yields: 
\begin{equation}
\exp \left[ -\frac{i}{\hbar}\int_{0}^{t}\frac{1}{2}m\omega^2 \left(x_{2o}+\int_{0}^{t}\omega(\tau)d\tau \right)^2 d\tau \right].
\end{equation}
This contains second degree in white noise which makes it difficult to deal with. To remedy this, we apply the Taylor series expansion as specified in ref. \cite{b.a}
\begin{eqnarray}
S_{V}(x_{2})&\approx& S_{V}(x_{2o}) + \frac{1}{1!}\int{}{}d\tau \omega(\tau) \frac{\partial S_{V}(x_{2o})}{\partial \omega(\tau)} 
\label{taylor}
\\
& & + \frac{1}{2!}\int{}{}d\tau_{1}d\tau_{2} \omega(\tau_{1}) \frac{\partial^2 S_{V}(x_{2o})}{\partial \omega(\tau_{1})\partial \omega(\tau_{2})}\omega(\tau_{1}). 
\nonumber
\end{eqnarray}
For simplicity we choose the initial point $x_{2o}=0$ which leads to $S_{V}(x_{2o})=0$ and
\begin{eqnarray}
S'&=&\frac{\partial S_{V}(0)}{\partial \omega(\tau)}=\frac{\hbar}{m}\int V'(0) d\tau \Rightarrow 0,  
\label{S'}
\\
S"&=&\frac{\partial^2 S_{V}(0)}{\partial \omega(\tau_{1})\partial \omega(\tau_{2})}=\frac{\hbar}{m}\int_{\tau_{1}\vee \tau_{2}}^{t}V''(0) d\tau 
\nonumber
\\
&\Rightarrow& \hbar \omega^2(t-\tau_{1} \vee \tau_{2}).
\label{S"}
\end{eqnarray}
Then, with Eqs. (\ref{donsker2}), (\ref{S'}) and (\ref{S"}) we can rewrite the propagator as
\begin{equation}
K_{x_{2}}=\int_{-\infty}^{+\infty}\frac{\exp \left[ -i\lambda x_{2} \right]}{2\pi}\left[ T.I\left(\sqrt{\frac{\hbar}{m}}\lambda \right) \right] d\lambda, 
\label{K1dim}
\end{equation}
where
\begin{equation}
I=N \exp \left[ -\frac{1}{2}\left\langle \omega, -(i+1)\omega \right\rangle \right] \exp \left[ -\frac{1}{2}\left\langle \omega, \frac{i}{\hbar}S"\omega \right\rangle \right],
\label{Iho}
\end{equation} 
and the evaluation of the Feynman path integral is carried by the T-transform \cite{b.a} given by $T.I(\xi=\sqrt{\frac{\hbar}{m}}\lambda) = \int_{}^{}I \exp \left[ i\left\langle \omega, \sqrt{\frac{\hbar}{m}}\lambda \right\rangle \right] d\mu(\omega)$ which can be simplified as
\begin{eqnarray}
T.I=\left[ det\left( 1+L(K+1)^{-1}\right)\right]^{-\frac{1}{2}} \exp \left[ -\frac{1}{2}(K+L+1)^{-1}\int_{0}^{t}\left( \sqrt{\frac{\hbar}{m}}\lambda \right)^2 d\tau \right], 
\label{T-transform}
\end{eqnarray}
where $K=-(i+1)$ and $L=i\hbar^{-1}S"$. Then, we substitute Eq. (\ref{T-transform}) into Eq. (\ref{K1dim}). Simplifying, we obtain
\begin{eqnarray}
K_{x_{2}}=\frac{1}{2\pi}\left[ det\left( 1+L(K+1)^{-1}\right)\right]^{-\frac{1}{2}}\int_{-\infty}^{+\infty}d\lambda \exp \left[ \frac{-\hbar t(K+L+1)^{-1}}{2m}\lambda^2-iq_{1} \lambda \right].
\label{whitenoisepropagatorK1dim2}
\end{eqnarray}
However, we note that $\left( 1+L(K+1)^{-1}\right)=(1-\hbar^{-1}S")$ and $(K+L+1)^{-1}=i(1-\hbar^{-1}S")^{-1}$. We can then rewrite Eq. (\ref{whitenoisepropagatorK1dim2}) as 
\begin{eqnarray}
K_{x_{2}}=\frac{1}{2\pi}\left[ det(1-\hbar^{-1}S")\right]^{-\frac{1}{2}}\int_{-\infty}^{+\infty}d\lambda \exp \left[ \frac{-i\hbar t(1-\hbar^{-1}S")^{-1}}{2m}\lambda^2-ix_{2} \lambda \right].
\label{whitenoisepropagatorK1dim2change}
\end{eqnarray}
Utilizing the Gaussian integral formula, we obtain
\begin{eqnarray}
K_{x_{2}}=\frac{1}{2\pi}\left[ det(1-\hbar^{-1}S")\right]^{-\frac{1}{2}} \sqrt{\frac{2\pi m}{i\hbar t \left\langle e, (1-\hbar^{-1}S")e \right\rangle}} \exp \left[ \frac{imx_{2}^{2}}{2\hbar t \left\langle e, (1-\hbar^{-1}S")e \right\rangle} \right],
\label{whitenoisepropagatorK1dim3}
\end{eqnarray}
where the unit vector $e$ is defined as $e=t^{-\frac{1}{2}}\chi_{[0,t]}$. Then, after some simplification \cite{b.a, c.d} we get 
\begin{eqnarray}
det(1-\hbar^{-1}S") &=& \cos \omega t,
\label{determinant}
\\ 
\left\langle e, (1-\hbar^{-1}S")e \right\rangle &=& \frac{1}{\omega t}\tan \omega t.
\label{diagonalization}
\end{eqnarray}
Finally, using Eqs. (\ref{determinant}) and (\ref{diagonalization}), we obtain the $x_{2}$-dimension propagator as
\begin{eqnarray}
K(x_{2},0;t,0)=\sqrt{\frac{m\omega}{2\pi i\hbar t \sin \omega t}} \exp \left[ \frac{im\omega}{2\hbar}x_{2}^2\cot \omega t \right].
\label{whitenoisepropagatorK1dimfinal}
\end{eqnarray}

\subsubsection*{The evaluation of the propagator corresponding to the Lagrangian $L_{2}$}

Notice that the Lagrangian $L_{2}$ is just similar to that of the $x_{2}$-dimension Lagrangian. Thus by following the same procedure in evaluating $x_{2}$-dimension propagator, we obtain the propagators as
\begin{eqnarray}
K(x_{4},0;t,0)=\sqrt{\frac{m\omega}{2\pi i\hbar t \sin \omega t}} \exp \left[ \frac{im\omega}{2\hbar}x_{4}^2\cot \omega t \right].
\label{whitenoisepropagatorK2dimfinal}
\end{eqnarray}

\subsubsection*{The explicit form of the propagator for the first degenerate normal mode frequency}
We can now solve for the $K_{x_{2}x_{4}}=K(x_{2},x_{4};x_{2o}=0,x_{4o}=0;t,0)$ propagator which is just the product of Eqs. (\ref{whitenoisepropagatorK1dimfinal}), (\ref{whitenoisepropagatorK2dimfinal}) given by 
\begin{eqnarray}
K(x_{2},x_{4};0,0;t,0)=\left[ \frac{m\omega}{2\pi i\hbar t \sin \omega t} \right] \exp \left[ \frac{im\omega}{2\hbar}(x_{2}^2 + x_{4}^2) \cot \omega t \right].
\label{kfull}
\end{eqnarray}

\subsubsection{Propagator for the dynamics described by the second degenerate normal mode frequency}
We observe that this eigenvalue is similar to the previous subsection and that $x_{1}=x_{4}=0$ and $x_{2}=-x_{3}$, hence the total Hamiltonian given in Eq. (\ref{totalH}) becomes, 
\begin{equation}
H = \frac{p_{2}^2}{2m} + \frac{1}{2}m\omega^2 x_{2}^2 + \frac{p_{3}^2}{2m} + \frac{1}{2}m\omega^2 x_{3}^2, 
\label{h2}
\end{equation}
where $\omega_{2}=\omega_{3}=\omega = \sqrt{\lambda}$. Following the same procedure in subsection (\ref{1steigen}), we obtain the full propagator as
\begin{eqnarray}
K(x_{2},x_{3};0,0;t,0)=\left[ \frac{m\omega}{2\pi i\hbar t \sin \omega t} \right] \exp \left[ \frac{im\omega}{2\hbar}(x_{2}^2 + x_{3}^2) \cot \omega t \right].
\label{kfullh2}
\end{eqnarray}

\subsubsection{Propagator for the dynamics described by the nondegenerate normal mode frequencies}
Though the third and fourth eigenvalues are different, since both are nondegenerate, they describe the case where all elements in the system and environment are oscillating. Then the total Hamiltonian given in Eq. (\ref{totalH}) yields, 
\begin{equation}
H = \frac{p_{1}^2}{2m} + \frac{1}{2}m\omega^2 x_{1}^2 + \frac{p_{2}^2}{2m} + \frac{1}{2}m\omega^2 x_{2}^2+ \frac{p_{3}^2}{2m} + \frac{1}{2}m\omega^2 x_{3}^2 + \frac{p_{4}^2}{2m} + \frac{1}{2}m\omega^2 x_{4}^2 + x_{1}C(x_{2}+x_{3}+x_{4}), 
\label{h3h4}
\end{equation}
where $\omega_{1}=\omega_{2}=\omega_{3}=\omega_{4}=\omega = \sqrt{\frac{\lambda}{m}}$. From it, we observe that $x_{2}=x_{3}=x_{4}$ oscillate at the same amplitude while $x_{1}$ is different. So we simplify Eq. (\ref{h3h4}) becomes, 
\begin{eqnarray}
H &=& \frac{p_{1}^2}{2m} + \frac{1}{2}m\omega^2 x_{1}^2 + 3\left[ \frac{p_{2}^2}{2m} + \frac{1}{2}m\omega^2 x_{2}^2 \right] + 3C~x_{1}x_{2},
\nonumber \\
&=& \frac{1}{2}m_{1}\dot{x}_{1}^2 + \frac{1}{2}m_{1}\omega^2 x_{1}^2 + \frac{1}{2}m_{2}\dot{x}_{2}^2 + \frac{1}{2}m_{2}\omega^2 x_{2}^2 + \lambda~x_{1}x_{2},
\label{h3h4simplified}
\end{eqnarray}
where we take $m_{1}=m$ and $m_{2}=3m$ and which we can identify to have the resemblance that of a coupled harmonic oscillator. Using the Hamilton's canonical transformation above, we obtain the Lagrangian to be 
\begin{eqnarray}
L = \frac{1}{2}m_{1}\dot{x}_{1}^2 - \frac{1}{2}m_{1}\omega^2 x_{1}^2 + \frac{1}{2}m_{2}\dot{x}_{2}^2 - \frac{1}{2}m_{2}\omega^2 x_{2}^2 - \lambda~x_{1}x_{2}.
\label{l3l4simplified}
\end{eqnarray}
Then we utilize a coordinate transformation \cite{c.c} which decouples ${x}_{1}$ and ${x}_{2}$ given by
\[
 \left( 
{\begin{array}{c}
  x_{1} \\
  x_{2}
\end{array}}
 \right) =  \left[ 
{\begin{array}{cc}
 (\frac{m}{m_{1}})^{\frac{1}{2}}\cos \phi & (\frac{m}{m_{1}})^{\frac{1}{2}}\sin \phi \\
 -(\frac{m}{m_{2}})^{\frac{1}{2}}\sin \phi & (\frac{m}{m_{2}})^{\frac{1}{2}}\cos \phi
\end{array}}
 \right] 
\left( 
{\begin{array}{c}
  y_{1} \\
  y_{2}
\end{array}}
 \right), 
\]
then after some algebra we can rewrite the Lagrangian in Eq. (\ref{l3l4simplified}) as
\begin{equation}
L = \frac{1}{2}m \left( \dot{y}_{1}^2 + \dot{y}_{2}^2 \right) - \alpha~y_{1}^2 - \beta~y_{2}^2 - \gamma ~y_{1}y_{2},
\end{equation}
where
\begin{eqnarray}
\alpha &=& \frac{1}{2}m\omega^2 - \frac{m\lambda}{2\sqrt{m_{1}m_{2}}}\sin 2\phi,
\label{aplha1}
\\
\beta &=& \frac{1}{2}m\omega^2 + \frac{m\lambda}{2\sqrt{m_{1}m_{2}}}\sin 2\phi,
\label{beta1}
\\
\gamma &=& \frac{m\lambda}{\sqrt{m_{1}m_{2}}}\cos 2\phi.
\label{gamma1}
\end{eqnarray}
To eliminate the system-bath coupling, $\gamma$ must vanish, so that Eq. (\ref{gamma1}) yields
\begin{equation}
\theta = \frac{(2n+1)\pi}{4},
\label{Dsolution}
\end{equation}
where $n=0,1,2,...$. Imposing the condition of Eq. (\ref{Dsolution}), we can finally write the Lagrangian in Eq. (\ref{l3l4simplified}) into $L=L_{3}+L_{4}$ with
\begin{eqnarray}
L_{3} &=& \frac{1}{2}m\dot{y}_{1}^2 - \frac{1}{2}m\Omega_{1}^2 y_{1}^2,
\label{x3}
\\
L_{4} &=& \frac{1}{2}m\dot{y}_{2}^2 - \frac{1}{2}m\Omega_{2}^2 y_{2}^2,
\label{x4}
\end{eqnarray}
where $\Omega_{1}^2=\omega^2 - \frac{\lambda}{\sqrt{m_{1}m_{2}}}$ and $\Omega_{2}^2=\omega^2 + \frac{\lambda}{\sqrt{m_{1}m_{2}}}$. Notice that Eqs. (\ref{x3}) and (\ref{x4}) are similar to that of Eqs. (\ref{l1newest}) and (\ref{l2newest}) in subsection (\ref{1steigen}), then by following the same evaluation procedure yields
\begin{eqnarray}
K_{x_{3}}&=&\left[ \frac{m\Omega_{1}}{2\pi i\hbar t \sin \Omega_{1} t} \right]^{\frac{1}{2}} \exp \left[ \frac{im\Omega_{1}}{2\hbar}y_{1}^2 \cot \Omega_{1} t \right],
\label{kx3} \\
K_{x_{4}}&=&\left[ \frac{m\Omega_{2}}{2\pi i\hbar t \sin \Omega_{2} t} \right]^{\frac{1}{2}} \exp \left[ \frac{im\Omega_{2}}{2\hbar}y_{2}^2 \cot \Omega_{2} t \right].
\end{eqnarray}
Furthermore we note that the coordinate transformation changes the path-integral measures ($D[x_{1}]D[x_{2}]=JD[y_{1}]D[y_{2}]$) through a Jacobian $J=\frac{\sqrt{m_{1}m_{2}}}{m}$, hence the propagator becomes
\begin{eqnarray}
K_{x_{3}x_{4}}=\frac{1}{2\pi i\hbar t}\left[ \frac{m_{1}m_{2}\Omega_{1}\Omega_{2}}{\sin \Omega_{1} t \sin \Omega_{2} t} \right]^{\frac{1}{2}} \exp \left[ \frac{im}{2\hbar}\left( y_{1}^2 \cot \Omega_{1} t + y_{2}^2 \cot \Omega_{2} t \right) \right].
\label{kx3x4} 
\end{eqnarray}
Then the desired propagator can be obtained by returning to the original coordinates
\[
 \left( 
{\begin{array}{c}
  y_{1} \\
  y_{2}
\end{array}}
 \right) =  \left[ 
{\begin{array}{cc}
 (\frac{m_{1}}{m})^{\frac{1}{2}}\cos \phi & -(\frac{m_{2}}{m})^{\frac{1}{2}}\sin \phi \\
 (\frac{m_{1}}{m})^{\frac{1}{2}}\sin \phi & (\frac{m_{2}}{m})^{\frac{1}{2}}\cos \phi
\end{array}}
 \right] 
\left( 
{\begin{array}{c}
  x_{1} \\
  x_{2}
\end{array}}
 \right), 
\]
thus with $m_{1}=m$ and $m_{2}=3m$ and doing some simplication on the propagator $K(x_{1},x_{2};0,0;t,0)=K_{x_{3}x_{4}}$, Eq. (\ref{kx3x4}) becomes 
\begin{eqnarray}
K_{x_{3}x_{4}}&=&\frac{m}{2\pi i\hbar t}\left[ \frac{3\sqrt{\omega^2 - \frac{\lambda}{\sqrt{m_{1}m_{2}}}}\sqrt{\omega^2 + \frac{\lambda}{\sqrt{m_{1}m_{2}}}}}{\sin \left( \sqrt{\omega^2 - \frac{\lambda}{\sqrt{m_{1}m_{2}}}} \right) t \sin \left( \sqrt{\omega^2 + \frac{\lambda}{\sqrt{m_{1}m_{2}}}} \right) t} \right]^{\frac{1}{2}} 
\label{kx3x4final} \\
&\times&\exp \left[ \frac{im}{2\hbar}\sqrt{\omega^2 - \frac{\lambda}{\sqrt{m_{1}m_{2}}}}\left( \left[ \frac{1}{2}x_{1}^2 - \sqrt{3}x_{1}x_{2} + \frac{3}{2}x_{2}^2 \right] \cot \left( \sqrt{\omega^2 - \frac{\lambda}{\sqrt{3}m}} \right) t \right) \right]
\nonumber \\
&\times&\exp \left[ \frac{im}{2\hbar}\sqrt{\omega^2 + \frac{\lambda}{\sqrt{m_{1}m_{2}}}}\left( \left[ \frac{1}{2}x_{1}^2 + \sqrt{3}x_{1}x_{2} + \frac{3}{2}x_{2}^2 \right] \cot \left( \sqrt{\omega^2 + \frac{\lambda}{\sqrt{3}m}} \right) t \right) \right].
\nonumber 
\end{eqnarray}

\subsubsection{The full propagator for the dynamics of the system}
Finally, we can obtain the entire system-environment propagator for $N=4$ by taking the  product of all propagators in Eqs. (\ref{kfull}), (\ref{kfullh2}) and (\ref{kx3x4final}). The resulting propagator,$K_{F}=K(x_{1},x_{2},x_{3},x_{4};0,0,0,0;t,0)$, has the following form:
\begin{eqnarray}
K_{F}&=&\left(\frac{m}{2\pi i\hbar t}\right)^3 \left(\frac{\omega}{\sin \omega t}\right)^2 \left[ \frac{3\sqrt{\omega^2 - \frac{\lambda}{\sqrt{3}m}}\sqrt{\omega^2 + \frac{\lambda}{\sqrt{3}m}}}{\sin \left( \sqrt{\omega^2 - \frac{\lambda}{\sqrt{3}m}} \right) t \sin \left( \sqrt{\omega^2 + \frac{\lambda}{\sqrt{3}m}} \right) t} \right]^{\frac{1}{2}} 
\label{kfullfinal} \\
&\times&\exp \left[ \frac{im}{2\hbar}\sqrt{\omega^2 - \frac{\lambda}{\sqrt{3}m}}\left( \left[ \frac{1}{2}x_{1}^2 - \sqrt{3}x_{1}x_{2} + \frac{3}{2}x_{2}^2 \right] \cot \left( \sqrt{\omega^2 - \frac{\lambda}{\sqrt{3}m}} \right) t \right) \right]
\nonumber \\
&\times&\exp \left[ \frac{im}{2\hbar}\sqrt{\omega^2 + \frac{\lambda}{\sqrt{3}m}}\left( \left[ \frac{1}{2}x_{1}^2 + \sqrt{3}x_{1}x_{2} + \frac{3}{2}x_{2}^2 \right] \cot \left( \sqrt{\omega^2 + \frac{\lambda}{\sqrt{3}m}} \right) t \right) \right]
\nonumber \\
&\times&\exp \left[ \frac{im\omega}{2\hbar}\left( \left[ 2x_{2}^2 + x_{3}^2 + x_{4}^2 \right] \cot \omega t \right) \right].
\nonumber
\end{eqnarray}

\subsection{Generalization to the case where there are $N$ oscillators in the environment}
As mentioned previously, generalizing our methods to the case where there are a large number $N$ of oscillators in the environment is not that difficult since its evaluation is closely related to that of the $N=4$ case. In fact, following the same procedure as that given in section (\ref{N4}) for $N$ oscillators in the environment and noting that there will always be two nondegenerate and $N-2$ degenerate normal mode frequencies, one can show that for large $N\geq 4$ the full propagator becomes 
\begin{eqnarray}
&&K_{F}=\left(\frac{m}{2\pi i\hbar t}\right)^{N-1} \left(\frac{\omega}{\sin \omega t}\right)^{N-2} \left[ \frac{(N-1)\sqrt{\omega^2 - \frac{\lambda}{\sqrt{N-1}m}}\sqrt{\omega^2 + \frac{\lambda}{\sqrt{N-1}m}}}{\sin \left( \sqrt{\omega^2 - \frac{\lambda}{\sqrt{N-1}m}} \right) t \sin \left( \sqrt{\omega^2 + \frac{\lambda}{\sqrt{N-1}m}} \right) t} \right]^{\frac{1}{2}} 
\label{kNfullfinal} \\
&\times&\exp \left[ \frac{im}{2\hbar}\sqrt{\omega^2 - \frac{\lambda}{\sqrt{N-1}m}}\left( \left[ \frac{1}{2}x_{1}^2 - \sqrt{N-1}x_{1}x_{2} + \frac{N-1}{2}x_{2}^2 \right] \cot \left( \sqrt{\omega^2 - \frac{\lambda}{\sqrt{N-1}m}} \right) t \right) \right]
\nonumber \\
&\times&\exp \left[ \frac{im}{2\hbar}\sqrt{\omega^2 + \frac{\lambda}{\sqrt{N-1}m}}\left( \left[ \frac{1}{2}x_{1}^2 + \sqrt{N-1}x_{1}x_{2} + \frac{N-1}{2}x_{2}^2 \right] \cot \left( \sqrt{\omega^2 + \frac{\lambda}{\sqrt{N-1}m}} \right) t \right) \right]
\nonumber \\
&\times&\exp \left[ \frac{im\omega}{2\hbar}\left( \left[ (N-2)x_{2}^2+\sum_{j=3}^{N}x_{j}^2 \right] \cot \omega t \right) \right].
\nonumber
\end{eqnarray}
Observe that the first two exponential expressions in the propagator correspond to the nondegenerate normal mode frequencies, while the last exponential expression in the propagator corresponds to the degenerate normal mode frequencies.


\section{Conclusion}
In this article, we have successfully solved for the quantum propagator of an open quantum system consisting of a simple harmonic oscillator which is coupled to an environment of N multimode. We first compute for the normal mode frequencies of the system, obtaining two non-degenerate and $N-2$ degenerate normal modes for $N \geq 4$ oscillators in the environment in order to decouple the coordinates of the system and the environment. 

The work done in this article shows that the method of white noise analysis posits promise in evaluating the propagators for open quantum systems, due to its mathematical rigor and ease of use. In particular, it can be applied to systems with N coupled oscillators which are all coupled to an environment, which can be used to model quantum transport of energy excitations in solid state and biological systems. The authors will explore these areas in further detail in future work.  

\begin{acknowledgments}
B. M. Butanas Jr would like to thank Prof. J. P. Esguerra and M. A. Pedroso-Butanas for the fruitful exchange of ideas and to the Philippines' Department of Science and Technology (DOST) and Central Mindanao University (CMU) for the scholarship and financial support via Accelerated Science and Technology Human Resource Development Program (ASTHRDP) and a faculty development program (FDP), respectively. R. C. F. Caballar would like to thank M. A. A. Estrella for conceptual discussions that clarified matters related to this work. B. M. Butanas Jr. and R. C. F. Caballar would like to thank National Institute of Physics, College of Science and UP Diliman for support and for providing a stimulating research atmosphere.
\end{acknowledgments}

\nocite{*}
\providecommand{\noopsort}[1]{}\providecommand{\singleletter}[1]{#1}%
\end{document}